\documentclass[11pt,twoside]{article}
\usepackage{./asp2014}
\usepackage{graphicx}
\usepackage{wrapfig}
\usepackage{lscape}
\usepackage{rotating}

\newcommand\nh{NH$_3$~}
\newcommand\kms{km s$^{-1}$}

\aspSuppressVolSlug
\resetcounters

\begin{document}

\title{Tracing the Water Snowline in Protoplanetary disks with the ngVLA}
\author{Ke Zhang$^1$, Edwin A. Bergin$^1$, Jonathan P. Williams$^2$, Paola Pinilla$^3$, Sean M. Andrews$^4$}
\affil{$^1$University of Michigan, Ann Arbor, MI 48105, USA; \email{kezhang@umich.edu}}
\affil{$^2$Institute for Astronomy, Honolulu, HI  96822, USA}
\affil{$^3$Department of Astronomy/Steward Observatory, The University of Arizona, Tucson, AZ 85721, USA}
\affil{$^4$Harvard-Smithsonian Center for Astrophysics, 60 Garden Street, Cambridge, MA 02138, USA}

\paperauthor{Ke Zhang}{kezhang@umich.edu}{0000-0002-0661-7517}{University of Michigan}{}{Ann Arbor}{MI}{48105}{USA}
\paperauthor{Edwin Bergin}{ebergin@umich.edu}{0000-0003-4179-6394}{University of Michigan}{}{Ann Arbor}{MI}{48105}{USA}
\paperauthor{Jonathan P. Williams}{jw@hawaii.edu}{0000-0001-5058-695X}{Institute for Astronomy}{}{Honolulu}{HI}{96822}{USA}
\paperauthor{Paola Pinilla}{pinilla@email.arizona.edu}{0000-0001-8764-1780}{Department of Astronomy/Steward Observatory}{}{Tucson}{AZ}{85721}{USA}
\paperauthor{Sean M. Andrews}{sandrews@cfa.harvard.edu}{0000-0003-2253-2270}{Harvard-Smithsonian Center for Astrophysics}{}{Cambridge}{MA}{02138}{USA}

\section{Introduction}

Water is believed to be the most crucial molecule for the habitability of planetary systems \citep{vanDishoeck14}. Its abundance and distribution in a protoplanetary disk greatly affect the atmospheric and core composition of giant planets \citep{Oberg11} and the surface water content of terrestrial planets \citep{Raymond07}. 

The most critical location in the water distribution is where water vapor condenses into ices, the so-called, \textit{water snowline}.  In the Solar Nebula,  the water snowline appears to be the critical boundary that separates the formation of terrestrial planets and giant planets \citep{Hayashi81}. 

Theoretical models predict the water snowline is a pivotal location for planetary architecture, as two important physical changes occur at this transition (see Figure~1). The first one is an enhancement in the surface density of solids--as water accounts for about 50\% of the total condensible mass, its condensation will significantly increase the solid mass beyond the snowline and accelerate the planet(esimal) formation \citep{Stevenson88, Pollack96}. The second change is the dust size distribution. Icy aggregates are significantly more resistant to compaction than bare silicate aggregates and thus icy aggregates can readily grow to a much larger size \citep{Ros13, Zhang15}. Once a critical decimeter-sized pebble population is formed, streaming instabilities can readily drive the creation of km-sized or larger planetesimals \citep{Johansen14}. On the other hand, the sintering effect may break aggregates and leads to a pile-up of smaller particles in the region slightly outside the snowline \citep{Okuzumi16}. Therefore the water snowline may play an important role in regulating the formation and chemical composition of planetesimals, and ultimately the formation of planets and their bulk composition.


\begin{figure}[]
\begin{center}
\includegraphics[scale=0.85,angle=0]{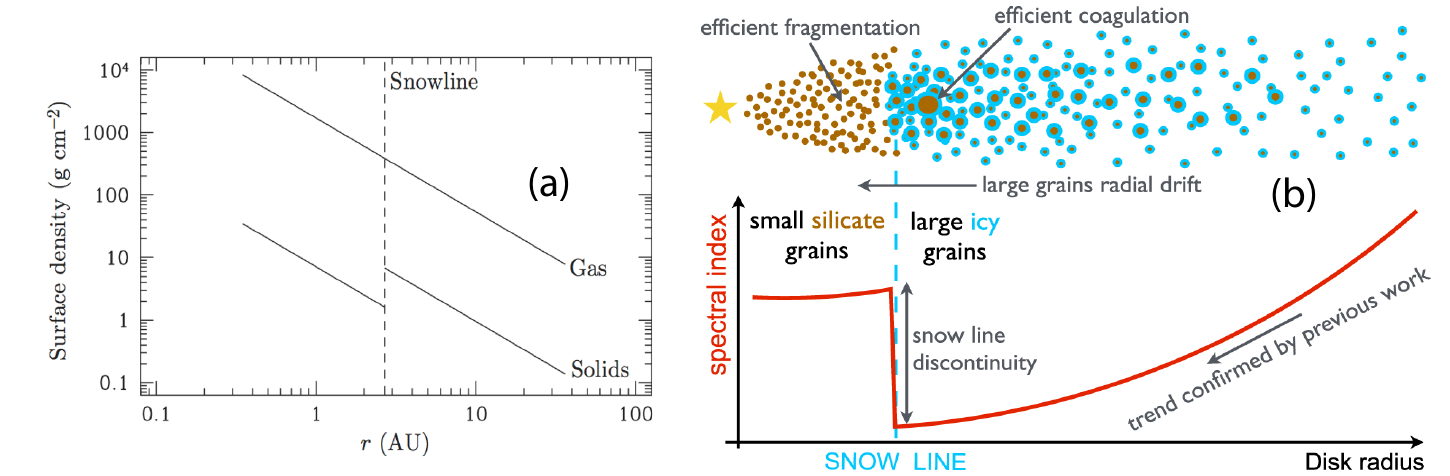}
\end{center}
\caption{\footnotesize {(a) The sharp discontinuity in the surface density of solids across the water snowline, based on Hayashi's minimum mass Solar Nebula model. The Figure is adapted from \citet{Armitage13}. (b) An illustrative example of the sharp discontinuity in spectral index $\alpha$ across the water snowline, as a result of the change in dust particle size distribution. The dust emission is assumed to be optically thin. The figure is from \citet{Banzatti15}.
}}
\end{figure}

Despite the importance of the water snowline, it is challenging to pinpoint its location and to quantitatively study its role in planetesimal formation. First of all, planetesimal formation occurs at the dust-rich mid-plane of the disk, while water forms quite readily in disk surface layers \citep{Bethell09}. Thus water line observations can only constrain the surface snowline \citep{Zhang13, Blevins16}.  Second, the water snowline in an accreting protoplanetary disk is around 1-10\,AU \citep{Kennedy08, Notsu16, Notsu17}. At these radii, the dust emission at relevant line frequencies of  water or its isotopologues (>=100\,GHz) is largely optically thick and thereby hindering a direct detection of the snowline. 

\section{Tracing the water snowline with ngVLA}
Alternatively, the water snowline maybe traced through the sharp changes of physical properties (surface density and dust size distribution) or through the snowlines of other molecules that sublimate co-spatially with water. In both approaches, observations are needed to be taken at frequencies where dust emission is optically thin and with a sufficient spatial resolution to resolve the region inside the snowline. The ngVLA is the only facility that can provide these capabilities.

\subsection{Tracing the water snowline through dust property changes}

A sharp change of dust properties across a water snowline may have been seen in recent ALMA observations of the V883 Ori system \citep{Cieza16}. V883 Ori is an FU Ori type source that is undergoing a large outburst in luminosity arising from a temporary increase in the accretion rate. The luminosity outburst warms the disk and pushes the water snowline to a large radius where the dust emission is less optically thick. The 230\,GHz continuum emission of the V883 Ori disk shows an intensity break due to an abrupt change in the optical depth at about 42\,AU, where the disk temperature is close to the water condensation temperature. Furthermore, the spectral index across the snowline is consistent with predictions from numerical simulations that include the coagulation, fragmentation and radial drift of dust grains \citep{Banzatti15}.  

The V883 Ori study was possible thanks to its large snowline during the luminosity outburst. For the majority of protoplanetary disks, their snowlines are only at 1-10\,AU, where the dust emission is highly optically thick at all ALMA bands. Furthermore, even in the V883 Ori case, the region inside the snowline (42\,AU) appears to be optically thick at 230\,GHz and thus hinders constraints on the dust size distribution.  Therefore, to study the dust size change across a snowline in a large number of disks, observations need to be carried out at much lower frequencies where the dust emission is optically thin. In addition, a higher spatial resolution is needed to spatially resolve the region inside the snowline. Only the proposed ngVLA facility has sufficient sensitivity and resolution to carry out this type of observations. 

To resolve a 5\,AU water snowline in a disk at a distance of 100\,pc, an angular resolution $\le$50\,mas is needed and a maximum recoverable scale of $\ge$500\,mas is needed to compare regions inside and outside the snowline. To meet the combination of resolution and low dust optical depth, the ideal wavelengths for this experiment is at 30\,GHz. The spectral index can be measured across a bandwidth of 20\,GHz. For a fiducial disk, the emission is predicted to be strong and a sensitivity of 5\,$\mu$Jy/beam would be sufficient. Higher frequency ALMA observations at a comparable spatial resolution can provide complementary measurements on the dust temperature through optically thick emission, which helps distinguish dust property changes caused by other mechanisms. Through these spectral index observations, we can study whether dust grains grow faster in regions beyond the water snowline and quantitatively constrain the size distribution and surface density of solids across the snowline. 

The ngVLA observations will provide the critical measurement of the water snowline locations in disks of different ages and stellar properties. The water snowline locations can be compared to snowlines of other abundant volatiles, such as the CO snowline measured by ALMA \citep{Qi13, Zhang17}. The snowlines of major volatiles set the rough boundaries of planetesimals with different elemental compositions, which further determine the bulk composition of terrestrial planets and cores of gas or ice giants \citep{Oberg11}.  The spectral index measurements will constrain the dust size distribution and surface density changes across the snowline. These will shed crucial insight into our understanding of the planetesimal formation and the birth locations of giant planets.

\subsection{Tracing the water snowline through a proxy: the NH$_3$ snowline}
Ammonia (NH$_3$) is a known major constituent of interstellar ices ($\sim$5\%, \citealt{Oberg11b}) and an important tracer of nitrogen chemistry in protoplanetary disks. 
Laboratory work of ice evaporation has shown at least a portion, or perhaps all of the NH$_3$ ice releases co-spatially with water \citep{Martin-Domenech14}. Furthermore, NH$_3$ has inversion transitions at cm wavelengths where the dust continuum emission is less optically thick. Given these two advantages, \nh is the best proxy candidate to trace the water snowline in protoplanetary disks. 

To resolve the \nh snowline, we can either spatially resolve the \nh emission inside its snowline or obtain high S/N spectra to constrain the gas emitting area through Keplerian velocity field in the disk as well as the centroid of the emission. As the lines are predicted to be faint (see below), the latter case is a more practical approach. 

\begin{figure}[]
\begin{center}
\label{fig:nh3}
\includegraphics[scale=0.3,angle=0]{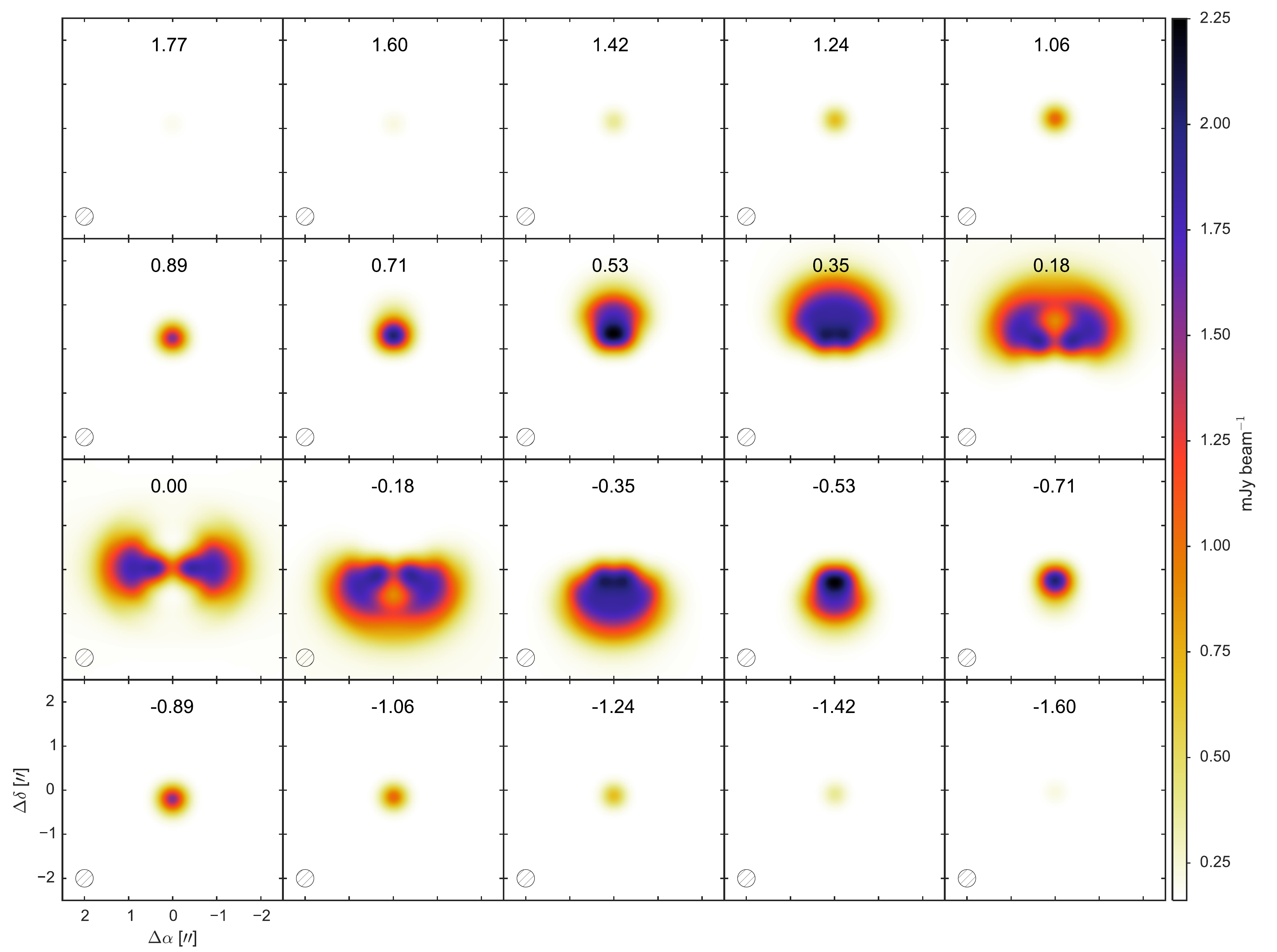}
\end{center}
\caption{\footnotesize {Emission maps as a function of velocity in a 0.\arcsec4 beam for the 1,1 transition of \nh in our fiducial disk model based on TW Hya \citep{Du15}.
}}
\end{figure}


 To explore detectability, we assume a minimal model -- one that sets limits for how faint \nh emission would be in the most stringent case.  This model assumes no \nh ice is provided from the ISM and \nh is made solely by the disk.   We then use the model of TW Hya that matches water vapor emission \citep{Du15} to predict emission in the 1,1 and 2,2 inversion transitions around 23\,GHz.  This predicted channel map emission from this model is shown in Figure~2) for TW Hya.  This needs to be scaled by one order of magnitude as most disk sources lie a factor of 3 more distant than TW Hya.  In the future, we can perform additional calculations that assume interstellar ices are present and where sublimation would raise the gas phase abundance and hence emission levels.
For a \nh emission detection, we find that a 10$\sigma$ detection of ammonia 1,1 transition towards TW Hya requires 0.25\,mJy in 0.2 km/s channels. To survey \nh reliably in hundreds of disk systems we need to be able to detect this line at 150 pc at the same level, requiring 25 $\mu$Jy in 0.2 km/s channels (10$\sigma$).  However, since lower velocity resolution can be used for detections, a sensitivity of 120\,$\mu$Jy in 1.0 km/s channels (5$\sigma$) would be sufficient for a survey.

The detection of the snowline via the model provided here appears be challenging as the required sensitivity in sources at 150\,pc would be 30 $\mu$Jy to detect the emission from the line wings at 150\,pc (using the model shown in Figure~2).  Since this represents a minimal model (a maximum \nh abundance of $\sim$ 2$\times$10$^{-6}$), any additional level of ice evaporation will increase the emission at these channels.   Furthermore, we likely will need to observe disks surrounding more massive stars such as B and A spectral type.  Here we still must use velocity line profiles to reach material at a few AU.  Thus, if \nh was present in the gas at the level of 5\% relative to water ice then its emission within the evaporated zone would be substantially increased, at least until the line becomes optically thick.  We estimate that the line will become optically thick when the line is about a factor of 10 higher than the current model or at 300 $\mu$Jy in 0.2\,\kms~channels.
\section{Conclusion}
The water snowline in protoplanetary disks is one of the most pivotal locations for planet formation: it not only sets a critical boundary of chemical composition in the disk, but may also serve as a favorable site of planetesimal growth and planet formation. The water snowline at the mid-plane cannot be directly traced by water line observations as the dust emission around the water snowline is highly optically thick at these line frequencies.

 In this chapter, we discuss two alternative approaches to trace the mid-plane water snowline -- by observing sharp transitions in dust properties across the snwoline or by using the \nh snowline as a proxy.  Either approach requires observations at cm wavelengths where the dust emission is likely to be optically thin. We found that with the current design of the ngVLA ability, a sharp transition in dust property across the water snowline can be readily traced through measuring the spectral index of continuum emissions of relatively bright disks. The \nh snowline detection is likely to be challenging under the current design, but the \nh abundance can be an order of magnitude higher than our conservative model and thus still be detectable. In summary, the proposed ngVLA is the only facility can provide sufficient sensitivity and spatial resolution to trace the evolution of water snowline at the disk mid-plane and reveal its role in the formation of planetesimals and in building the architecture of planetary systems.





\end{document}